\documentclass{article}

     \PassOptionsToPackage{numbers}{natbib}

\usepackage[final]{neurips_2023}

\usepackage[utf8]{inputenc} 
\usepackage[T1]{fontenc}    
\usepackage{hyperref}       
\usepackage{url}            
\usepackage{booktabs}       
\usepackage{amsfonts}       
\usepackage{nicefrac}       
\usepackage{microtype}      
\usepackage{xcolor}         
\usepackage{setspace}
\usepackage{amsmath}
\usepackage{amssymb} 
\usepackage{amsthm}
\usepackage{amsfonts, mathtools}
\usepackage{algorithm,algpseudocode}
\usepackage{bm}
\usepackage{bbm}
\usepackage{comment}
\usepackage{graphicx}
\usepackage{multirow, booktabs}
\usepackage{caption}
\usepackage{subcaption}

\usepackage{siunitx} 
\usepackage{adjustbox}

\title{Facilitating Battery Swapping Services for Freight Trucks with Spatial-Temporal Demand Prediction}

\author{%
  Linyu Liu$^{1,3}$ \  \ \ Zhen Dai$^{2}$ \ \ \ Shiji Song$^{3}$ \ \ \ Xiaocheng Li$^{4}$ \ \ \ Guanting Chen$^{1}$ \thanks{Corresponding to: guanting@unc.edu} \\
  $^{1}$ Department of Statistics and Operations Research, UNC-Chapel Hill\\
  $^{2}$ Chongqing Expressway Group Company, Chongqing, China\\
  $^{3}$ Department of Automation, Tsinghua University\\
  $^{4}$ Imperial College Business School, Imperial College London\\
}

\begin{document}

\maketitle

\begin{abstract}
Electrifying heavy-duty trucks offers a substantial opportunity to curtail carbon emissions, advancing toward a carbon-neutral future. However, the inherent challenges of limited battery energy and the sheer weight of heavy-duty trucks lead to reduced mileage and prolonged charging durations. Consequently, battery-swapping services emerge as an attractive solution for these trucks. This paper employs a two-fold approach to investigate the potential and enhance the efficacy of such services. Firstly, spatial-temporal demand prediction models are adopted to predict the traffic patterns for the upcoming hours. Subsequently, the prediction guides an optimization module for efficient battery allocation and deployment. Analyzing the heavy-duty truck data on a highway network spanning over 2,500 miles, our model and analysis underscore the value of prediction/machine learning in facilitating future decision-makings. In particular, we find that the initial phase of implementing battery-swapping services favors mobile battery-swapping stations, but as the system matures, fixed-location stations are preferred.
\end{abstract}



\section{Introduction}

The global challenge of carbon emissions has exhibited different patterns and witnessed various responses across sectors. In 2021, 21.2\% of these emissions are attributed to the transportation sector (\cite{bouckaert2021net}). While substantial progress has been made in the electrification of passenger vehicles, heavy-duty trucks remain notably falling behind in this transition. In contrast, heavy-duty trucks, though accounting for less than 10\% of the total vehicle population, are responsible for more than 40\% of carbon dioxide emissions (\cite{hao2019impact, moultak2017transitioning, xue2019toward}). This disparity underscores the opportunity for mitigating transportation environment impact via the electrification of heavy-duty trucks. In practice, several factors contribute to the lagging electrification of heavy-duty trucks, chiefly among them are the limitations in battery size and energy density, range anxiety, and long charging durations. However, as EV technology advances and battery costs decrease, battery swapping stations are emerging as an economically viable solution for business proprietors and station operators (\cite{zhu2023does}). 

To take China as an example, state-owned highway companies have shown significant interest in exploring the commercial potential of establishing battery swapping stations (BSS) specifically for heavy-duty trucks. These entities, responsible for constructing and overseeing state-owned highways and associated infrastructure (such as service areas, power stations, and charging stations), posit that (i) service areas spaced every 50 miles, (ii) combined with swift battery swapping services, and (iii) electric vehicles' reduced operational cost per mile (compared with traditional vehicles), can strongly boost the electrification for heavy-duty trucks.

\begin{figure}[t]
\centering
\begin{subfigure}[r]{0.5\textwidth}
\includegraphics[width = \linewidth]{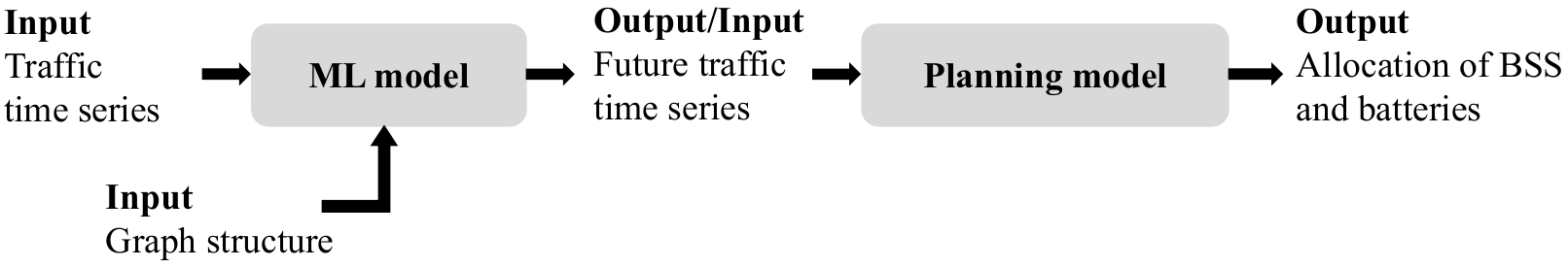}
\caption{Our methodology: The ML model takes two inputs of traffic time series and road network/graph and outputs the future traffic prediction for a window of the next few hours. Then the prediction is fed into the planning/optimization model which decides the final battery allocation of fixed and mobile BSS.}
\label{fig:method} 
\end{subfigure}
\hfill
\begin{subfigure}[l]{0.45\textwidth}
\includegraphics[width = \linewidth]{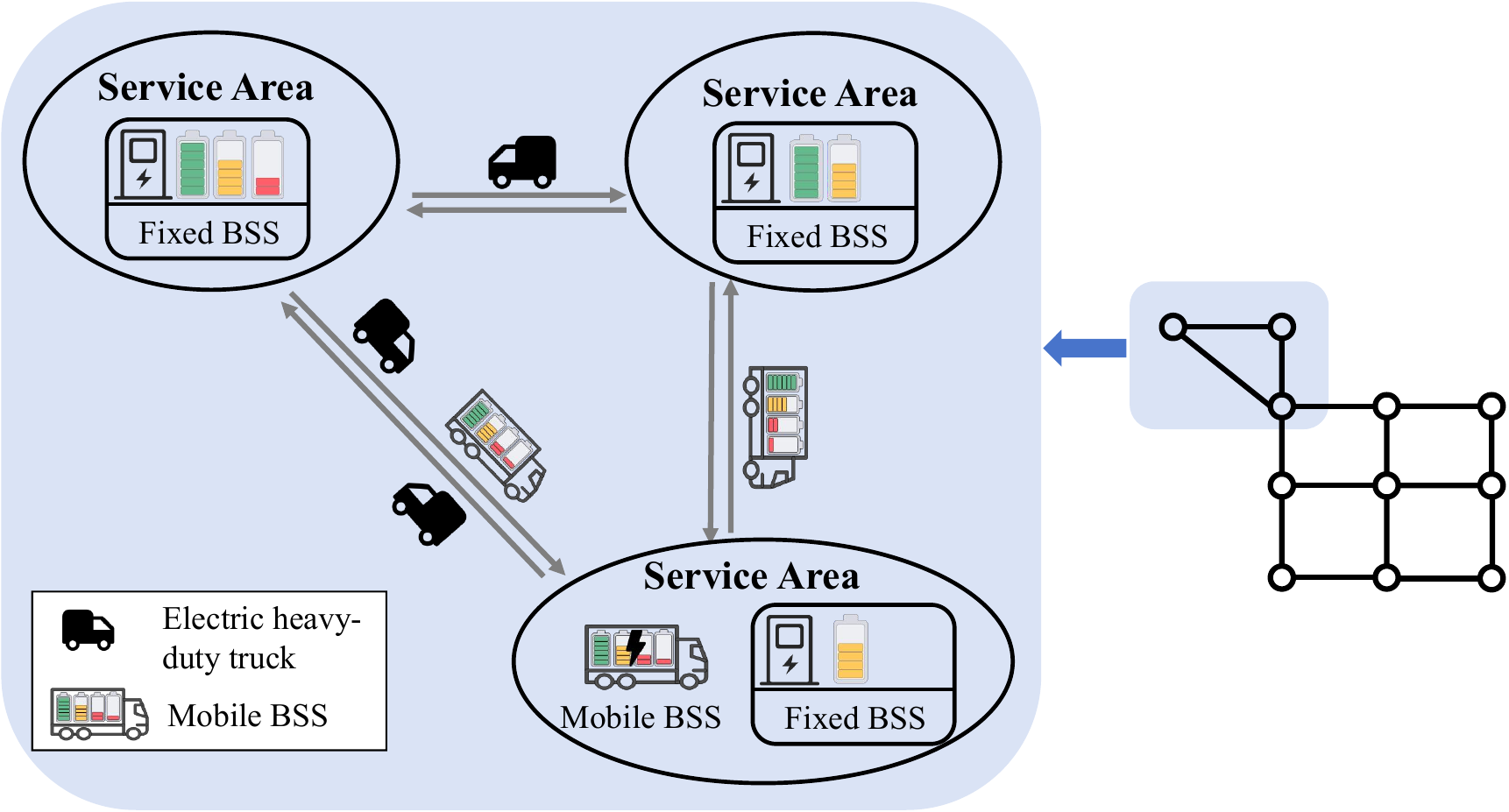}
   \caption{The battery allocation model and network}
   \label{fig:network} 
\end{subfigure}
\end{figure}

This work presents a data-driven approach to optimize the operational efficiency of such BSS services. Our methodology adopts a ``predict-then-optimize'' pipeline (Figure \ref{fig:method}). Utilizing traffic data of heavy-duty trucks, we employ spatial-temporal machine learning prediction models to forecast traffic volume in upcoming hours. Leveraging these predictions, we then solve an integer program to derive a dynamic scheduling policy for BSS operations. Our work is in collaboration with the Chongqing Expressway Group Company (CEGC), a key entity in charge of the development and management of highways in Chongqing, China, a municipality spanning 31,700 square miles (the largest municipality area in the world) and a population over 32 million.

Our prediction approach is closely related to the literature on machine learning methods for traffic forecasting. As neural network (NN) methods evolve (\cite{hochreiter1997long, krizhevsky2012imagenet,kipf2016semi,vaswani2017attention}), NN-based traffic prediction methods (\cite{park1999forecasting,fu2016using,zhao2019t,wang2020traffic}) have become increasing effective. In the traffic prediction literature, there is also a growing interest in leveraging machine methods for climate-change-related goals (\cite{blattner2021commercial,rollend2022machine,buechler2021evgen,kapoor2020helping}). The predicted traffic volume is used to optimize the operations of BSS, which is related to the vast literature of planning and management for electric vehicles  (\cite{mak2013infrastructure,widrick2018optimal,koirala2022planning,qi2023scaling,sarker2014optimal}) and energy systems (\cite{yang2022modelling,powell2022charging,borlaug2021heavy}).

Here we summarize our key findings. From a prediction perspective, based on our dataset, we observe that the integration of the attention module (\cite{vaswani2017attention}) enhances the prediction accuracy of the GCN-based model for longer horizons. In terms of the service rate,  compared to the hindsight optimal strategy where the ground truth is known, our predict-then-optimize approach demonstrates robust performance across various environments. Our investigation also sheds light on the planning choice between fixed BSS (a BSS with a fixed location) and mobile BSS (a BSS that can change its location). We observe that the traffic pattern significantly influences decision-making for infrastructure choices more than the accuracy of traffic prediction: in the initial phases of implementing battery-swapping services, the traffic has a more irregular pattern/heteroskedasticity, which favors the mobile BSS; as the system matures, the traffic has lower variation and more stable pattern, and in this case, the fixed BSS is more preferred.

\section{Data and Methodology}\label{sec_model}

\subsection{Data}

In collaboration with CEGC, we analyze traffic data for heavy-duty trucks on highways. The traffic data is captured by cameras mounted on gantries across an expansive network that stretches over 2,500 miles. Positioned roughly every 2 miles, there are more than 1,800 of these structures dotting the network. As vehicles pass beneath these gantries, the cameras record the precise time and take images. CEGC subsequently processes these images using computer vision techniques (analyzing the vehicle's appearance and its license plate) and thus identifies the vehicle type.

We construct a network topology using data collected over a span of three months (Figure \ref{fig:network}), where nodes represent service areas designated for the launch of fixed BSS. These service areas also serve as locations where mobile BSS units operate. Meanwhile, the edges of our topology represent the highway routes, facilitating the movement of both heavy-duty trucks and mobile BSS.

\subsection{Model}
\textbf{Prediction.} We employ machine learning models to forecast traffic for heavy-duty trucks on highways. We focus on short-term predictions and target a window length of the next 6 hours, which is the time scale for the operations of BSS. The predicted traffic flow then serves as the input to optimize the operations of BSS. Specifically, we denote $\mathcal{D}_t = \{\bm{x}_1,\bm{x}_2, \cdots, \bm{x}_t\}$ the traffic data up to time $t$, where $\bm{x}_t \in \mathbb{R}^m$ denote the cumulative traffic on $m$ edges during the time interval $(t-1, t]$. Denoting $f$ the machine learning model, and $h$ the prediction interval, we intend to train $f$ such that the output $f(\mathcal{D}_t) = [\hat{\bm{y}}_1, \hat{\bm{y}}_2,\cdots,\hat{\bm{y}}_h]^\top$ is close to the observed future traffic $[\bm{x}_{t+1},\cdots,\bm{x}_{t+h}]^{\top}$. We choose the following machine learning models for $f$ and aim to investigate the effect of the self-attention mechanism on traffic prediction tasks.
\begin{itemize}
\item 
\textbf{Temporal Graph Convolutional Network (T-GCN)} (\citep{zhao2019t}): T-GCN combines the graph convolutional network (GCN) and the gated recurrent unit (GRU) to incorporate the spatial and temporal information to forecast traffic volumes.
\item \textbf{Attention Graph Convolutional Network (A3T-GCN)}(\citep{bai2021a3t}): A3T-GCN modifies the T-GCN by incorporating an attention module to dynamically capture the spatial and temporal correlations of traffic volumes.
\end{itemize}

We report the performance of these two methods in Table \ref{table:ml_perf}.



\textbf{Planning/Optimization.} In our setting, the fixed BSS encompasses a battery-swapping booth along with an adjacent inventory house dedicated to charging and storing batteries. The mobile BSS is a sizable truck equipped with a battery-swapping module and storage module (the storage module is also capable of charging). Since the traffic of heavy-duty trucks is highly non-stationary, the mobile BSS offers such flexibility when the demand at certain locations surges. By carefully dispatching spare batteries between service areas, mobile BSS can effectively match the temporally and spatially varying demands for battery swaps. 

Now we describe a scheduling optimization model that dispatches these mobile batteries among service areas. In our highway network, the travel time for heavy-duty trucks between adjacent battery swapping stations (approximately 40 miles) is approximately 1 hour. As of the year 2023, one 282 kWh battery for heavy-duty trucks also takes roughly 1 hour to get fully charged with an ultra-fast charger (\citep{saadaoui2023overview}). Henceforth, we discretize time into hourly intervals, with a unit time step representing 1 hour. The optimization model needs to decide whether each mobile BSS should remain at the current station or be dispatched and relocated to a nearby station. If the latter option is adopted, the model needs to further decide the number of batteries to dispatch. The moving battery on the way cannot be used for service until it arrives at the destination. 

In our optimization model, the objective is to minimize the total lost demand over the horizon. At time $t$, the decision is made based on $[\bm{x}_1,\cdots,\bm{x}_t,\hat{\bm{y}}_1, \cdots, \hat{\bm{y}}_h]^{T}$. Although the algorithm looks into the future and makes multi-step planning,  we re-optimize the algorithm at every time step based on the new information $\mathcal{D}_{t+1}$ and the prediction $f(\mathcal{D}_{t+1})$. We refer to Appendix \ref{ap_opt} for details of the formulation and definitions.

We note that in this work, we assume all the trucks are electric trucks for simplicity but without loss of generality. Note that the goal of our analysis is to provide guidance for infrastructure planners to facilitate the electrification process, and this assumption helps circumvent the classic chicken-and-egg dilemma: full electrification demands robust infrastructure, but investing heavily in such infrastructure is not feasible unless there is a significant presence of electric trucks. However, the acquisition of these electric trucks presents a hefty upfront cost for logistic business owners. 

\section{Results}\label{sec_exp}

\textbf{Prediction accuracy.} 
Table \ref{table:ml_perf} compares the forecasting accuracy of T-GCN and A3T-GCN on the test set. T-GCN performs better for traffic prediction for the next $2$ hours, and A3T-GCN performs better for traffic prediction for the next 3-5 hours. Our hypothesis is that the self-attention module makes the future prediction sequences more adaptive, resulting in better performance in a longer forecasting window. A visualization of our predictions against actual traffic is presented in Figure \ref{fig_graph_acc}.

\begin{figure}[t]
\includegraphics[width = \linewidth]{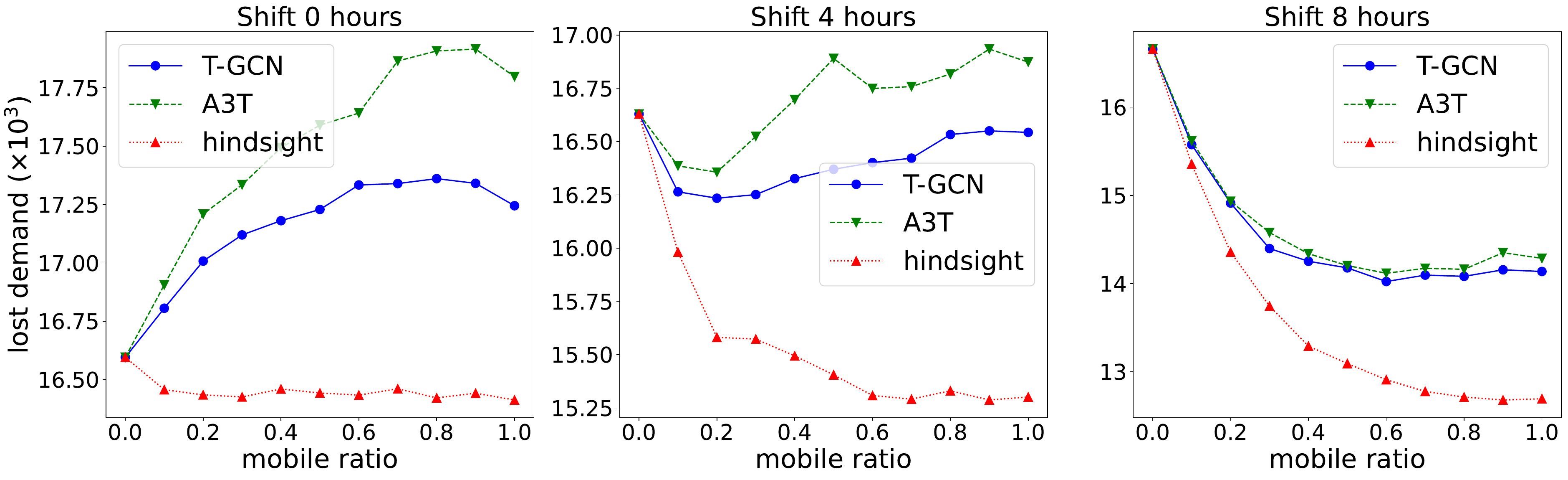}
\caption{Lost demand under different shifting hours.}
\label{fig:mana_ins}
\end{figure}

\textbf{Prediction-based optimization.} In Figure \ref{fig_varying_inventory}, we compute the lost demand on the test set, under the setting where the ratio of total batteries from fixed BSS v.s. from mobile BSS is $7:3$. We report the performance of two ML-based scheduling policies when the total inventory of batteries is high (90\% of the average), medium (75\% of the average), and low (60\% of the average). We find that these two methods perform similarly: both ML approaches feature a lost demand that is within $109\%$ of the lost demand of the oracle policy (the one that solves the optimization problem in a hindsight manner with the actual future traffic). The scheduling policy based on T-GCN consistently outperforms the A3T-GCN-based policy with a margin of around $2\%$ across all inventory levels. Our hypothesis is that compared with the longer prediction length $(h > 2)$, T-GCN features better accuracy for the shorter prediction length $(h \leq 2)$, which is more helpful for the predict-then-optimize pipeline.

In the same environment as above, we then investigate the effect of the prediction length $h$ on the final performance of the scheduling policy. The scheduling policy (except for the oracle policy) takes $\{\bm{x}_1,\cdots,\bm{x}_t,\hat{\bm{y}}_1,\cdots,\hat{\bm{y}}_h\}$ as input to make decision, and we vary $h$ to see the effect on the lost demand. According to Figure \ref{fig:diff_planning_len_bar}, we find that for all scheduling policies (including the oracle policy), the larger the $h$, the lower the lost demand. This indicates that although the prediction on longer horizon is less accurate, it has positive effect on improving the performance of the scheduling policy. Lastly, see Figure \ref{fig:visualization-shift} for visualization for the scheduling policy where mobile BSS tries to catch demand.

\textbf{Fixed BSS vs mobile BSS.} Lastly, we discuss the choice between deploying fixed BSS and mobile BSS. CEGC intends to launch the battery swapping service on selected highway routes, targeting contracts with logistics companies that own truck fleets operating between industrial areas. 
As a result, during the early stages, electric truck's traffic volume is likely to exhibit peaks that are distinct from the peaks of the cumulative traffic volume of trucks. This ushers in a different pattern to our existing traffic dataset. To simulate these nonstationary structural patterns in the early stage of electrification, we randomly select various nodes (service stations), and for their adjacent edges (routes), we advance the traffic volume along the time axis, ensuring that traffic peaks are altered (see Figure \ref{fig_early_bss}).

In Figure \ref{fig:mana_ins}, we evaluate the lost demand of the T-GCN-based scheduling model under different fixed BSS to mobile BSS ratios and different degrees of traffic shift. We find that when there is no traffic shift (this corresponds to the situation with full electrification of heavy-duty trucks), the oracle policy improves moderately as the ratio of mobile BSS increases. However, the prediction error of ML hinders the performance of both ML-based scheduling policies, resulting an increased lost demand. When there is a large amount of traffic shift (this corresponds to the early stage of electrification), both ML-based scheduling policies show a considerable amount of increases in performance. The managerial insight is that two factors favor the adoption of mobile BSS: (i) irregular patterns of the traffic volume and (ii) high accuracy of the demand prediction model. If there is a large improvement in the prediction accuracy, mobile BSS is very effective even when the full electrification of heavy-duty trucks happens.

\newpage 

\section*{Acknowledgments and Disclosure of Funding}
This work utilizes traffic data provided by Chongqing Expressway Group Company. The authors would like to thank Mengxiong Zhou, Jiahong Li, and Jie Yang for their collaboration on this project.

This work was partially supported by the National Natural Science Foundation of China under Grant 61936009, and partially by the National Science and Technology Innovation 2030 Major Project of the Ministry of Science and Technology of China (NO.2018AAA0101604).

\bibliographystyle{plainnat} 
\bibliography{sample.bib} 

\newpage

\appendix

\section{Details for the Optimization Model}\label{ap_opt}

We schedule these mobile batteries in a rolling manner: in each time step, we solve the optimization problem to obtain a scheduling plan for the next $T$ hours, but only the scheduling plan for the current (first) time step is executed. We assume that there is a sufficient number of trucks serving as mobile BSS, ensuring that at every time step, we can deliver any battery in the mobile BSS to any nearby stations. We outline our model in model \eqref{equ:planning_model} associated with notations defined in Table \ref{tab:notations}.

\begin{subequations}
\begin{align}
\min_{\bm{L},\bm{z}} \ & \sum_{t=1}^T\sum_{i\in\mathcal{S}} L_{t,i} \label{equ:obj}\\ 
{\rm s.t.}\ & \sum_{j\in\mathcal{S}} z_{1,i,j} = Q_i, \ \ \forall i\in\mathcal{S} \label{equ:cons-1}\\
& \sum_{j\in\mathcal{S}}z_{t-1,j,i} = \sum_{j\in\mathcal{S}} z_{t,i,j}, \ \ \forall t=2,...,T,\ \forall i\in\mathcal{S} \label{equ:cons-2}\\
& z_{t,i,j} = 0, \ \ \forall t=1,...,T, \ \forall j\in\mathcal{S}\backslash \mathcal{N}(i),\ \forall i\in\mathcal{S} \label{equ:cons-3}\\
& L_{t,i} \ge D_{i,t} - (F_i+z_{t,i,i}), \ \ \forall t=1,...,T-1,\ \forall i\in\mathcal{S} \label{equ:cons-4}\\
& L_{T,i} \ge D_{i,t} - (F_i+\sum_{j\in\mathcal{S}} z_{T-1,j,i}), \ \ \forall i\in\mathcal{S} \label{equ:cons-5}\\
& z_{t,i,j}\in \mathbb{N}_+, \ \ \forall t=1,...,T,\ \forall i,j\in\mathcal{S}\\
& L_{t,i}\in \mathbb{N}_+, \ \ \forall t=1,...,T,\ \forall i\in\mathcal{S} \label{equ:cons-6}
\end{align}
\label{equ:planning_model}
\end{subequations}

\begin{table}[htbp]
    \centering
    \caption{List of Notations}
    \begin{tabular}{l p{0.7\linewidth}}
        \toprule
        \textbf{Notation} & \textbf{Description} \\
        \midrule
        \underline{Sets} & \\
        $\mathcal{S}$ & the set of service stations/areas\\
        $\mathcal{N}(i)$ & the neighboring stations of station $i$\\
        \underline{Parameters:} & \\
        $T$ & the planning horizon \\
        $Q_i$ & the initial number of mobile batteries at station $i$ \\
        $F_i$ & the number of fixed batteries at station $i$ \\
        $D_{i,t}$ & battery demand (prediction) at station $t$ at time step $t$ \\
        \underline{Decision Variables:} & \\
        $z_{t,i,j}$ & the number of batteries moving from station/area $i$ to $j$ at time step $t$ \\
        $L_{t,i}$ & the number of lost/unmet demand at station $i$ at time step $t$ \\
        \bottomrule
    \end{tabular}
    \label{tab:notations}
\end{table}

Constraints \eqref{equ:cons-1} and \eqref{equ:cons-2} are flow conservation constraints for the mobile batteries among stations. Constraint \eqref{equ:cons-3} only allows mobile BSS to transfer to adjacent stations in consecutive time steps. Constraints \eqref{equ:cons-4}, \eqref{equ:cons-5}, and \eqref{equ:cons-6} describe the definition of lost demand. The total number of batteries equals inventory level $\times$ average demand of all stations. The ratio of the total number of mobile/movable batteries and fixed/immovable batteries is set as the given mobile-to-fix ratio. At each station, the number of fixed batteries $F_i$ is set proportional to the average demand of station $i$. In the experiment, the initial value of $Q_i$ is determined by solving an optimization model similar to \eqref{equ:planning_model} with mobile battery total capacity constraint, which is formally written as

\begin{subequations}
\begin{align}
\min_{\bm{L},\bm{z},Q_i} \ & \sum_{t=1}^T\sum_{i\in\mathcal{S}} L_{t,i} \\
{\rm s.t.}\ & \sum_{i\in\mathcal{S}} Q_i = Q\\
& Q_i \in \mathbb{N}_+,\ \forall i\in\mathcal{S}\\
& \text{imposing constraints} \,\,\eqref{equ:cons-1}-\eqref{equ:cons-6} 
\end{align}
\end{subequations}
where $Q$ is the total number of mobile batteries.

\section{Other Graphs}
\begin{figure}[h]
\centering
\begin{subfigure}[c]{0.38\textwidth}
\begin{adjustbox}{width=\linewidth}
\begin{tabular}{|c|c|c|c|}
\hline
Hour & Algorithm & RMSE & MAE \\
\hline
\multirow{3}{*}{$1^{\rm st}$} & T-GCN&3.66 & 2.25 \\
& A3T-GCN&4.43 & 2.58 \\
\hline
\multirow{3}{*}{$2^{\rm nd}$} & T-GCN&4.16 & 2.41 \\
& A3T-GCN&4.48 & 2.53 \\
\hline
\multirow{3}{*}{$3^{\rm rd}$} & T-GCN&4.65 & 2.59 \\
& A3T-GCN&4.51 & 2.54 \\
\hline
\multirow{3}{*}{$4^{\rm th}$} & T-GCN&4.95 & 2.72 \\
& A3T-GCN&4.51 & 2.56 \\
\hline
\multirow{3}{*}{$5^{\rm th}$} & T-GCN&5.15 & 2.85 \\
& A3T-GCN&4.56 & 2.64 \\
\hline
\end{tabular}
\end{adjustbox}
\caption{Prediction performance}
\label{table:ml_perf}
\end{subfigure}
\hfill
\begin{subfigure}[c]{0.49\textwidth}
\includegraphics[width = \linewidth]{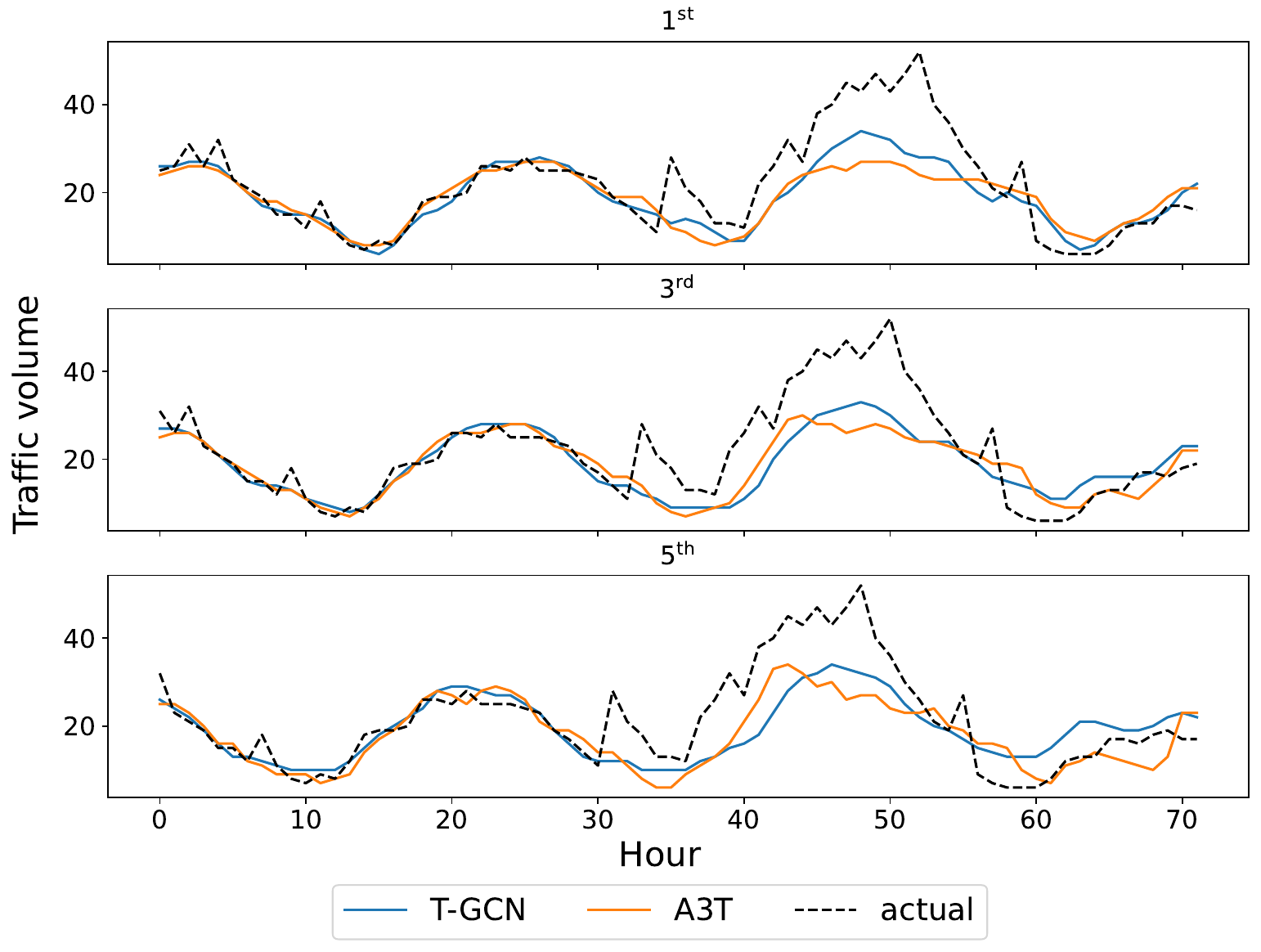}
   \caption{Visualization of prediction}
   \label{fig_graph_acc} 
\end{subfigure}
\caption{Table \ref{table:ml_perf} depicts the performance measure of T-GCN and A3T-GCN. The column Hour stands for the $\hat{\bm{y}}_s$ such that $s \in [1, h]$. We report performance metrics such as root mean squared error (RMSE) and mean absolute error (MAE). Figure \ref{fig_graph_acc} shows the predicted traffic volume compared with the actual traffic volume in a service station.}
\end{figure}

\begin{figure}[h]
\centering
\begin{subfigure}[c]{0.42\textwidth}
    \includegraphics[width=\linewidth]{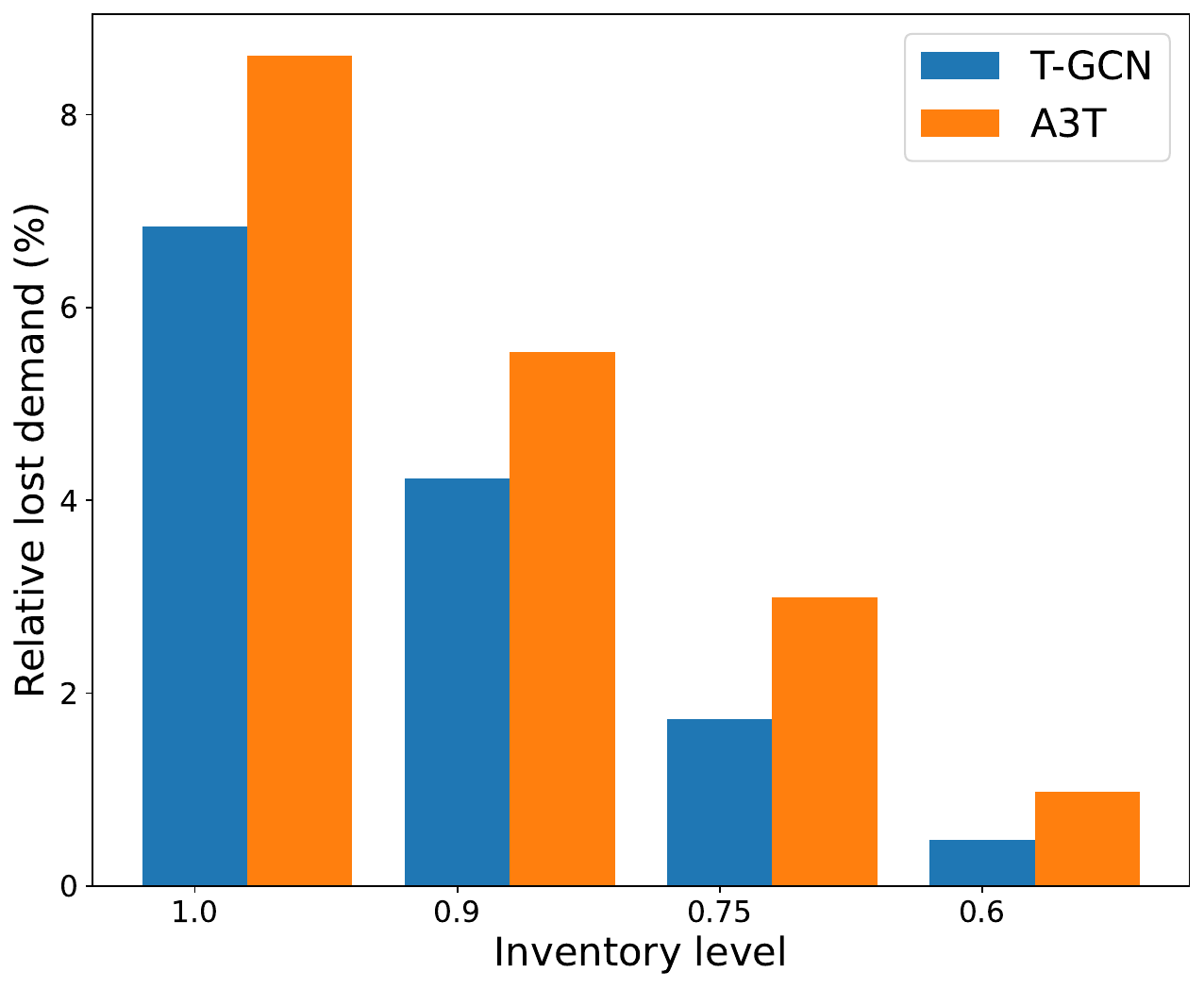}
    \caption{Lost demand under different inventory}
    \label{fig_varying_inventory}
\end{subfigure}
\hfill
\begin{subfigure}[c]{0.44\textwidth}

\includegraphics[width=\linewidth]{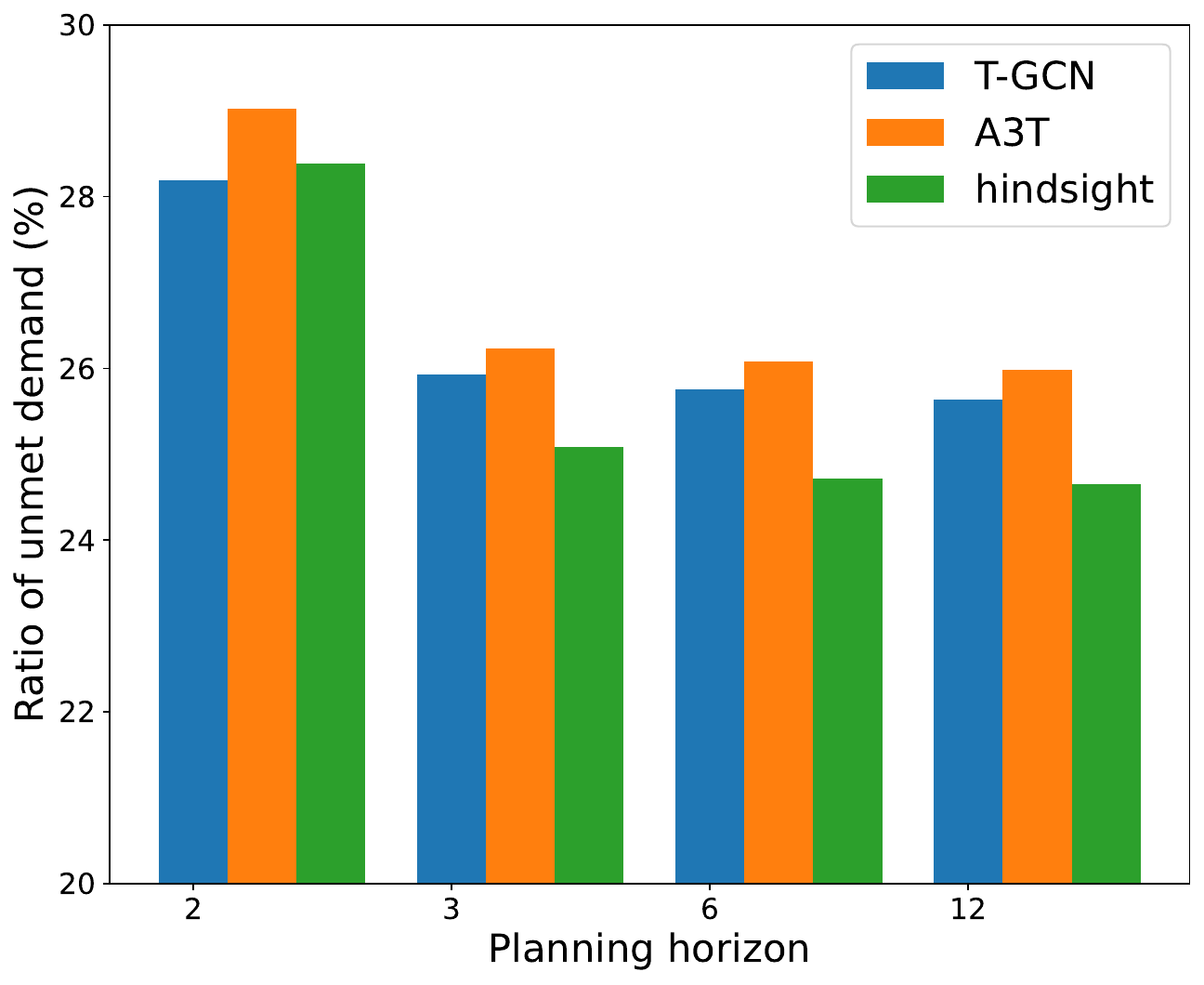}
\caption{Performance under different planning horizons}
\label{fig:diff_planning_len_bar}
\end{subfigure}
\caption{For \ref{fig_varying_inventory}, the horizontal axis stands for the ratio of the total inventory of batteries to the average demand, and the vertical axis stands for the relative lost demand compared to that of the oracle (hindsight) scheduling policy. For \ref{fig:diff_planning_len_bar}, the horizontal axis stands for the value of $h$. Based on different $h$, the scheduling policy outputs different unmet demand, and the vertical axis depicts the ratio of unmet demand to the total demand.}
\end{figure}

\begin{figure}[h]
\centering
\includegraphics[width = 0.5\linewidth]{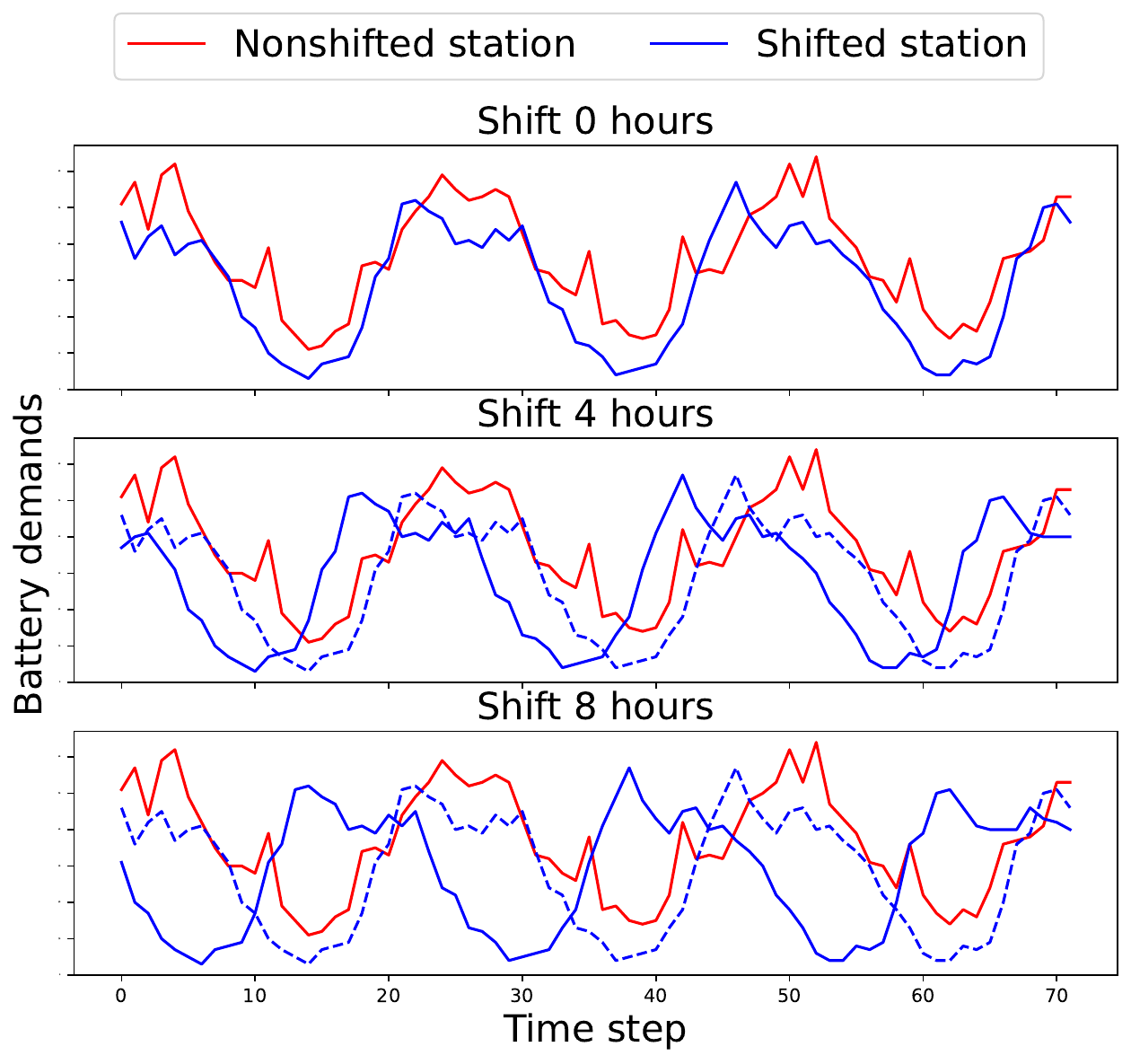}
\caption{Illustration of shifting. The red line represents the actual traffic of a service station whose traffic volume is not shifted. The blue lie represents a shifted one. The dotted line represents the traffic after shifting forward (4 hours and 8 hours). The traffic pattern of the road network will change after such a shift.}
\label{fig_early_bss} 
\end{figure}

\begin{figure}[h]
\centering
\includegraphics[width = \textwidth]{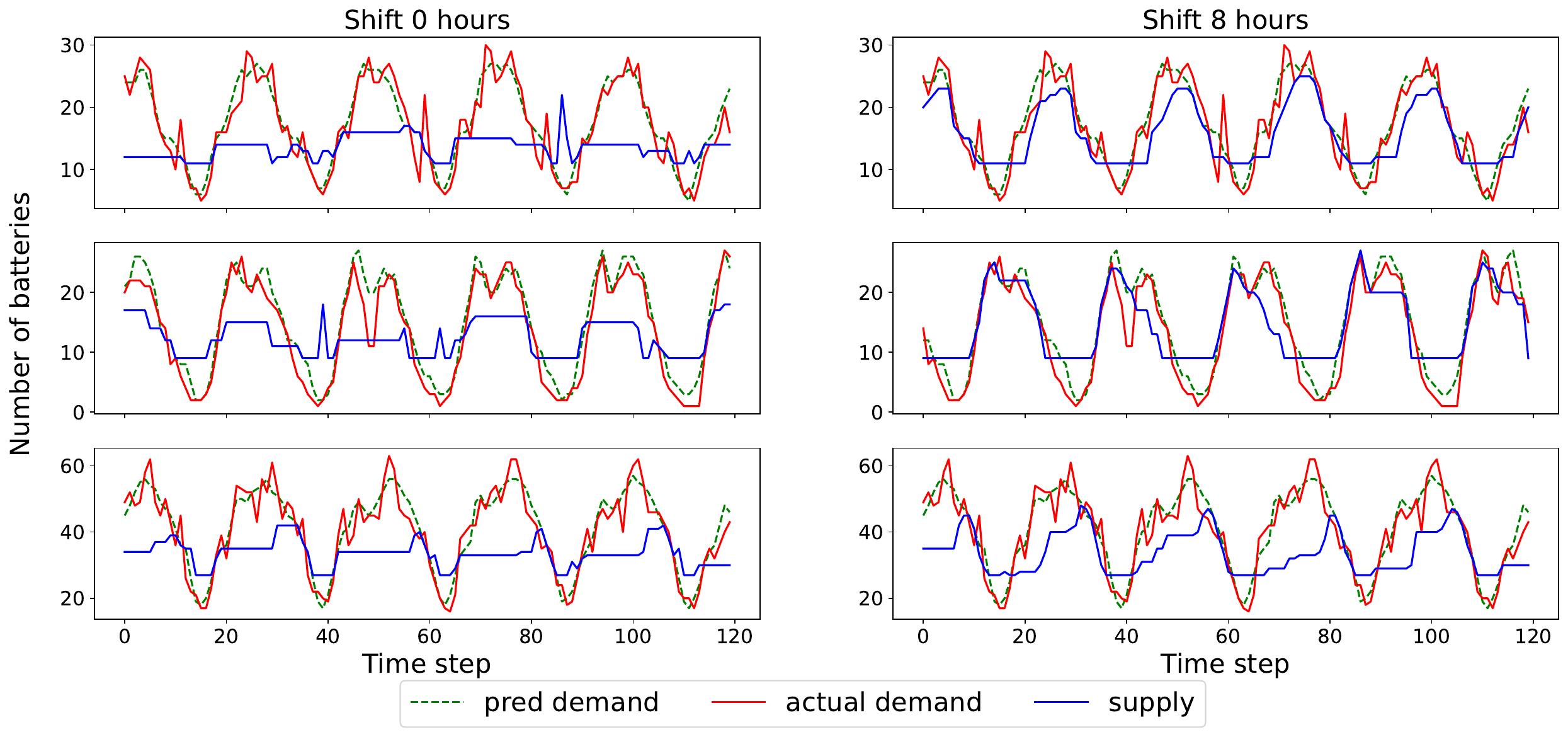}
\caption{Demand-supply visualization of 3 stations under mobile BSS to fixed BSS ratio $=0.3$. The green dotted line denotes the ML-predicted demand, the red line stands for the actual demand, and the blue lines shows the supply of the BSS controlled by the scheduling policy. From the plots, we can observe that the mobile BSS is allocated prior to the traffic surge, which demonstrates the effectiveness of the planning policy.}
\label{fig:visualization-shift}
\end{figure}

\end{document}